\documentclass{article}
\usepackage{spconf,amsmath,amssymb,amsfonts,graphicx,color,bbold,bbm}

\newtheorem{theorem}{Theorem}

\newtheorem{corollary}{Corollary}

\newtheorem{assumption}{Assumption}

\newcommand{\bs}{\boldsymbol}
\newcommand{\y}{\mathbf{y}}

\newcommand{\X}{\mathbf{X}}

\def \A {\mathbf{A}}

\def \C {\mathbf{C}}

\def \h {\mathbf{h}}
\def \H {\mathbf{H}}
\def \I {\mathbf{I}}

\def \S {\mathbf{S}}

\def \u {\mathbf{u}}
\def \v {\mathbf{v}}

\def \X {\mathbf{X}}

\def \y {\mathbf{y}}
\def \Z {\mathbf{Z}}
\def \z {\mathbf{z}}

\def \Hcal {\mathcal{H}}

\def \Kcal {\mathcal{K}}

\def \Ncal {\mathcal{N}}
\def \Ocal {\mathcal{O}}

\def \Vcal {\mathcal{V}}

\def \Cbb {\mathbb{C}}
\def \Ebb {\mathbb{E}}

\def \Pbb {\mathbb{P}}

\def \Zbb {\mathbb{Z}}

\def \drm {\mathrm{d}}

\def \erm {\mathrm{e}}
\def \irm {\mathrm{i}}

\def \epsilonbs {\boldsymbol{\epsilon}}

\def \xibs {\boldsymbol{\xi}}

\def \Sigmabs {\boldsymbol{\Sigma}}

\def \Xibs {\boldsymbol{\Xi}}

\def \diag{\mathrm{diag}}

\DeclareMathOperator*{\argmax}{argmax}

\title{On the frequency domain detection of high dimensional time series}
\name
{
	A. Rosuel$^{(1)}$, P. Vallet$^{(2)}$, P. Loubaton$^{(1)}$, X. Mestre $^{(3)}$
	\thanks{The authors were supported by the grant ANR-17-CE40-0003 of the French National Research Agency ANR (project HIDITSA).}
}
\address
{
	\normalsize
	$^{(1)}$ Laboratoire IGM (CNRS, Univ. Paris-Est/MLV), 5 Boulevard Descartes, 77454 Marne-la-Vall\'ee, France	
	\\
	\normalsize
	$^{(2)}$ Laboratoire IMS (CNRS, Univ. Bordeaux, Bordeaux INP), 351 Cours de la Lib\'eration, 33400 Talence, France
	\\
	\normalsize
	$^{(3)}$ CTTC, Av. Carl Friedrich Gauss 08860 Castelldefels, Barcelona, Spain
	\\
	\footnotesize\textsf{\{philippe.loubaton,alexis.rosuel\}@u-pem.fr, pascal.vallet@bordeaux-inp.fr, xavier.mestre@cttc.cat}
}

\begin{document}
\ninept
\maketitle
\begin{abstract}
	In this paper, we address the problem of detection, in the frequency domain, of a $M$-dimensional time series modeled as the output of a $M \times K$ MIMO 
	filter driven by a $K$-dimensional Gaussian white noise, and disturbed by an additive $M$-dimensional Gaussian colored noise. 
	We consider the study of test statistics based of the Spectral Coherence Matrix (SCM) obtained as renormalization of the smoothed periodogram matrix of the
	observed time series over $N$ samples, and with smoothing span $B$. 
	To that purpose, we consider the asymptotic regime in which $M,B,N$ all converge to infinity at certain specific rates, while $K$ remains fixed.
	We prove that the SCM may be approximated in operator norm by a correlated Wishart matrix, for which Random Matrix Theory (RMT) provides a precise description
	of the asymptotic behaviour of the eigenvalues. These results are then exploited to study the consistency of a test based on the largest eigenvalue of the SCM,
	and provide some numerical illustrations to evaluate the statistical performance of such a test.
\end{abstract}
\begin{keywords}
Spectral analysis, Detection Tests, High Dimensional Statistics, Random Matrix Theory
\end{keywords}
\section{Introduction}
\label{sec:intro}

The detection of a low rank multivariate signal corrupted by a spatially uncorrelated noise with unknown statistics is an important signal processing problem that is met in the context of array processing of uncalibrated sensor networks, see e.g. \cite{Boonstra2003} for applications to radio-astronomy, 
\cite{Ramirez2011} and \cite{SalaAlvarez2016a} motivated by detection of primary signals in the context of cognitive radio. 
We also notice that the underlying signal model, called the \emph{errors in variables model}, plays an important role in other fields such as 
econometrics, and was intensively studied in the past (see e.g. \cite{Soederstroem2007} for a review).

References \cite{Ramirez2011} and \cite{SalaAlvarez2016a} developed GLRT tests which require solving rather difficult optimization problems to be implemented, in particular if the underlying errors in variables model is \emph{dynamic} in the sense that the useful signal coincides with the output of a MIMO filter driven by a low-dimensional white noise and that the noise on each sensor is correlated in time (see \cite[Sec. V]{Ramirez2011}). 

The goal of this paper is to address the detection of the useful signal in the context of a dynamic errors in variables model when the observations dimension $M$ is large, the number of available samples $N$ is limited, and the rank $K$ of the spectral density of the useful signal is much smaller
than $M$ and $N$. This timely context is in particular motivated by the considerable development of large sensor networks which tend to produce high-dimensional multivariate signals. While the high dimensionality of the observations poses a number of new statistical problems, it sometimes allows to simplify the performance analysis of traditional statistical inference schemes as shown in the present paper. 

An important class of high-dimensional models, called the \emph{generalized dynamic linear factor models}, were introduced at the end of the nineties in econometrics, see e.g. \cite{Forni2000,Forni2004}. It is assumed that $M$ and $N$ both converge towards $+\infty$ in such a way that $\frac{M}{N}$ remains bounded ($\frac{M}{N}$ may also converges towards $0$) and that $K$ remains fixed. The fundamental assumption formulated in these works is that the $K$ non zero eigenvalues of the spectral density matrix of the useful signal converge towards $+\infty$ while the spectral density of the noise remains bounded. 
Under this regime, the $K$ largest eigenvalues of an estimate of the spectral density of the observation converge towards $+\infty$ almost surely, thus leading to a consistent detection scheme. Moreover, it is possible to retrieve the  frequency domain principal components of the useful signal.
While this regime is certainly justified in the econometrics field, it is not the most relevant in the context of high dimensional array processing where relevant algorithms have the potential to produce a Signal to Noise Ratio (SNR) ratio gain of the order of the number of sensors. 
Therefore, regimes in which the SNR $\rho$, before applying such algorithms, is $\mathcal{O}(\frac{1}{M})$ term, are of special interest. 
Unfortunately, the generalized dynamic linear factor models assumptions lead to a larger order of magnitude SNR.
 
In contrast, large random matrix methods allow to obtain interesting results in situations where $\rho = \mathcal{O}(\frac{1}{M})$. 
Previous related works addressed \emph{static models} (also known as \emph{narrowband models} in array processing) where a number of papers used the so-called spiked large random matrices defined as the sum of a low rank matrix due to the signal with a full rank random matrix representing the additive noise (see e.g. \cite{Benaych-Georges2012,Bianchi2011,Nadakuditi2008,Vallet2015,Vinogradova2013}).
The most complete results were obtained when the additive noise is temporally and spatially white. It appears that the standard detection test comparing 
the largest eigenvalue of the sample covariance matrix to a threshold is consistent if and only if $M \rho$ 
\footnote{If $K=1$, $M \rho$ represents the signal to noise after spatial matched filtering.}
is larger than the ratio $\sqrt{\frac{M}{N}}$. 
In other words, the high-dimensionality of the observations produces a threshold effect on standard detectors. 
While the existing large random matrix results allow to address detection in static models, a considerable work is still needed to consider \emph{dynamic models} (also known as \emph{wideband models} in array processing) in regimes where $\rho = \mathcal{O}(\frac{1}{M})$. 

In this paper, we propose to develop frequency domain methods where the sample covariance matrix is replaced by a frequency smoothed estimate $\hat{\C}(\nu)$ of the spectral coherence matrix of the observation at a Fourier frequency $\nu$. Under some reasonable assumptions, we establish that at each Fourier frequency $\nu$, $\hat{\C}(\nu)$ behaves as the sample covariance matrix of a static spiked large random matrix model whose noise part is temporally and spatially white. 
Using this useful result, we characterize the conditions under which the test comparing $\max_{\nu} \lambda_{max}(\hat{\C}(\nu))$ to a certain threshold is consistent. 

\emph{General notations}. Vectors and matrices are denoted respectively as bold lower case and bold upper case. If ${\bf A}$ is a matrix, $\| {\bf A} \|$ represents its spectral norm and $\| {\bf A} \|_{F}$ its Frobenius norm. If $a_1,\ldots,a_M \in \Cbb$, $\diag(a_1,\ldots,a_M)$ represents the diagonal matrix with diagonal elements $a_1,\ldots,a_M$, whereas if $\A$ is a matrix, $\diag(\A)$ is the diagonal matrix representing the diagonal part of $\A$. For any $M \times M$ Hermitian matrix $\A$, $\lambda_1(\A)\geq\ldots\geq\lambda_M(\A)$ denote the eigenvalues of $\A$ sorted in decreasing order. The notation $\Ncal_{\Cbb^M}(\mathbf{0},\I)$ is used for the standard $M$-dimensional complex Gaussian distribution.

%If ${\bf A}$ is reduced to a vector ${\bf a}$, $\| {\bf a} \|$ is equal to its Euclidian norm. 

\section{Model and assumptions}
\label{sec:problem}

In this section, we introduce the signal model and some definitions and assumptions that will be used throughout the next sections.

We consider a $M$--dimensional observed signal $(\y_n)_{n \in \mathbb{Z}}$ modeled as 
\begin{equation}
    {\bf y}_n={\bf u}_n+{\bf v}_n
    \label{definition:model}
\end{equation}
where $({\bf u}_n)_{n \in \mathbb{Z}}$ is a useful signal defined as the output of an unknown causal and stable $M \times K$ MIMO filter driven by a 
$\Ncal_{\Cbb^K}(\mathbf{0},\I)$ white noise $({\bs \epsilon}_n)_{n \in \mathbb{Z}}$, that is almost surely (a.s.),
\begin{align}
	\u_n = \sum_{k=0}^{+\infty} \H_k \epsilonbs_{n-k}
	\notag
\end{align}
%${\bf y}_n = $ 
%where ${\bf H}(\nu)=\sum_{k\ge0}{\bf H}_k e^{-2i\pi k\nu}$ represents the transfer function of the filter. $
and where the additive noise $\left(\v_n\right)_{n \in \Zbb}$ is modeled as a $M$--dimensional stationary complex Gaussian time series whose components time series $(v_{1,n})_{n \in \Zbb},\ldots,(v_{M,n})_{n \in \Zbb}$ are mutually independent. For each $\nu \in [0,1]$, we define 
$\H(\nu) = \sum_{k = 0}^{+\infty} \H_k \erm^{-\irm 2 \pi \nu k}$, and for each $m \in \{1,\ldots,M\}$,
we denote by $(r_m(k))_{k \in \Zbb}$ the covariance sequence of $(v_{m,n})_{n \in \Zbb}$ and by 
$s_m(\nu)$ the corresponding spectral density. 

%For each $\nu \in [0,1]$, we define
%\begin{align}
%	\H(\nu) = \sum_{k = 0}^{+\infty} \H_k \erm^{-\irm 2 \pi \nu k}
%	\notag
%\end{align}
%and for each $m \in \{1,\ldots,M\}$, we assume that the covariance sequence of $(v_{m,n})_{n \in \Zbb}$ denoted as $(r_m(k))_{k \in \Zbb}$ is summable, so that we have the representation
%\begin{align}
%	r_m(k) 
%	= \mathbb{E}\left[v_{m,n+k} v_{m,n}^{*}\right]
%	= \int_0^1 s_m(\nu) e^{2 \irm \pi k \nu} \drm\nu.
%	\notag
%\end{align}
%where $s_m$ denotes the corresponding spectral density.

Denote by
\begin{align}
	\xibs_{\y}(\nu) = \frac{1}{\sqrt{N}}\sum_{n=1}^N \y_n \erm^{-\irm 2 \pi \nu (n-1)}
	\notag
\end{align}
the finite Fourier transform of $(\y_n)_{n \in \Zbb}$ over the sample window $n=1,\ldots,N$, and for an even integer $B < N$ referred to as smoothing span,
\begin{align}
	\hat{\S}_{\y}(\nu) = \frac{1}{B+1} \sum_{b=-B/2}^{B/2} \xibs_{\y}(\nu + b/N) \xibs_{\y}(\nu + b/N)^*
	\notag
\end{align}
the classical frequency smoothed estimate of the spectral density $\S_{\y}(\nu)$ of $(\y_n)_{n \in \Zbb}$, where
\begin{align}
	\S_{\y}(\nu) = \H(\nu)\H(\nu)^* + \S_{\v}(\nu)
	\notag
\end{align}
with $\S_{\v}(\nu) = \diag(s_1(\nu),\ldots,s_M(\nu))$ the spectral density of $(\v_n)_{n \in \Zbb}$.
We recall (see e.g. the classical reference \cite{Brillinger1981}) that in the classical large sample size regime where $B,N\to \infty$ while $M,K$ are fixed,
$\Ebb[\hat{\S}_{\y}(\nu)] = \S_{\y}(\nu) + \Ocal(\frac{B^2}{N^2})$ and 
$\Ebb[\|\hat{\S}_{\y}(\nu)-\Ebb[\hat{\S}_{\y}(\nu)]\|^2] = \Ocal(\frac{1}{B})$ and the bias-variance compromise is achieved by choosing $B \to \infty$
such that $\frac{B}{N} \to 0$.

In the following, we study the statistical behaviour of the Spectral Coherence Matrix (SCM)
\begin{align}
	\hat{\C}_{\y}(\nu) = \diag\left(\hat{\S}_{\y}(\nu)\right)^{-\frac{1}{2}} \hat{\S}_{\y}(\nu) \ \diag\left(\hat{\S}_{\y}(\nu)\right)^{-\frac{1}{2}}
	\label{definition:C}
\end{align}
under the following high-dimensional asymptotic regime. We assume that $M = M(N)$ and $B = B(N)$ are both functions of $N$ such that 
\begin{align}
	M \sim N^{\alpha} \quad\text{and}\quad \frac{M}{B} \xrightarrow[N\to\infty]{} c \in (0,1)
	\label{assumption:regime}
\end{align}
for $\alpha \in (0,1)$, while $K$ remains fixed with respect to $N$. Choosing $B$ in such a way that $M/B \rightarrow 0$ ($B = \mathcal{O}(N^{\beta}), \beta > \alpha$) would make $\hat{\S}_{\y}(\nu)$ a consistent estimator of $\S_{\y}(\nu)$. However, in practice, 
for finite values of $M$ and $N$, it is not necessarily possible to choose $B$ in such a way
that $M \ll B$ and $B \ll N$, thus making the regime $M/B \rightarrow 0$ inaccurate. Assumption \ref{assumption:regime} appears thus relevant.

%Note that this asymptotic regime makes sense in practical situations where $M$ and $B$ are of the same order of magnitude 
%and much larger than $K$, and where $N$ is also much larger than $M,N$. 
%The condition $B \sim N^{\alpha}$, slightly stronger than $\frac{B}{N} \to 0$, is essentially due to the compensation of the regime $M \to \infty$ assumed here.

%\textcolor{red}{[Discuss the usual interpretation and compromises regarding $B$ and $N$]}.

Under this asymptotic regime, it is necessary to precise how certain quantities previously defined evolve with respect to $M$ (and thus $N$).
We first consider the following assumption regarding the noise part $(\v_n)_{n \in \Zbb}$.
%the short memory property of time series $(\u_n)_{n \in \Zbb}$ and $(\v_n)_{n \in \Zbb}$.
\begin{assumption}
	\label{assumption:noise}
	The following hold
	\begin{align}
		\limsup_{M \to \infty} \max_{m=1,\ldots,M} \sum_{k \in \Zbb} \left(1+|k|^2\right) |r_m(k)| < \infty
		\label{assumption:r}
	\end{align}
	and
	\begin{align}
		\liminf_{M \to \infty} \min_{m=1,\ldots,M} \inf_{\nu \in [0,1]} s_m(\nu) > 0
		\label{assumption:Snu}
	\end{align}
%	\begin{align}
%		\limsup_{M \to \infty}  \sum_{k =0}^{+\infty} \left(1+k\right) \left\|\H_k\right\| < \infty
%		\label{assumption:H}
%	\end{align}
\end{assumption}
Note that condition \eqref{assumption:r} trivially implies that the noise spectral densities $s_1,\ldots,s_M$ are twice continuously differentiable and that
\begin{align}
	\limsup_{M \to \infty} \max_{m=1, \ldots, M} \sup_{\nu \in [0,1]} |s_{m}^{(i)}(\nu)| < +\infty
	\notag
\end{align}
for $i \in \{0,1,2\}$, with $s_m^{(i)}$ the $i$-th order derivative of $s_m$.  
The condition \eqref{assumption:Snu} is intuitively expected as our objective is to study the behaviour of the spectral coherence matrix \eqref{definition:C}, which involves a renormalization by the inverse of the spectral density estimates of each time series (noise whitening).
Note also that condition \eqref{assumption:Snu} together with \eqref{assumption:r} implies that the total power of the noise satisfies
\begin{align}
	0 < \liminf_{M \to \infty} \frac{1}{M}\Ebb \left\|\v_n\right\|^2 \leq \limsup_{M \to \infty} \frac{1}{M}\Ebb \left\|\v_n\right\|^2 < \infty
	\notag
\end{align}
The next assumption is related to the signal part $(\u_n)_{n \in \Zbb}$.
%spectral behaviour of $(\u_n)_{n \in \Zbb}$ and $(\v_n)_{n \in \Zbb}$.
\begin{assumption}
	\label{assumption:signal}
	The following hold
	\begin{align}
		\limsup_{M \to \infty}  \sum_{k =0}^{+\infty} \left(1+k\right) \left\|\H_k\right\| < \infty
		\label{assumption:H}
	\end{align}
	and if $\h_1(\nu),\ldots,\h_M(\nu)$ denote the rows of $\H(\nu)$, then
	\begin{align}
		\lim_{M \to \infty}  \max_{m=1,\ldots,M} \sup_{\nu \in [0,1]} \left\|\h_m(\nu)\right\| = 0
		\label{assumption:Hnu}
	\end{align}
\end{assumption}
Since $K$ is assumed fixed with respect to $N$, condition \eqref{assumption:H} implies that $\Ebb \|\u_n\|^2 = \sum_{k \geq 0} \|\H_k\|_F^2 = \Ocal(1)$ so that the "observed SNR" vanishes as $M\to\infty$ at rate $1/M$, that is
%\begin{align}
	$\rho = \frac{\Ebb \|\u_n\|^2}{\Ebb \|\v_n\|^2} = \Ocal\left(\frac{1}{M}\right)$.
	%\notag
%\end{align} 
We notice that this regime is in accordance with a number of works studying e.g. the behaviour of large sensor array processing techniques \cite{Vallet2015}: taking advantage of large number $M$ of sensors to increase the observed SNR (or SNR before matched filtering in this context) $\rho$ by a factor $\Ocal(M)$, it is reasonable to expect that despite a low SNR, reliable information can be extracted on the useful signal.
In order to explain the significance of condition \eqref{assumption:Hnu}, we notice that the power $\mathbb{E}|u_{m,n}|^2$ of the contribution of $\u_n$ 
on sensor $m$ can be written as  $\mathbb{E}|u_{m,n}|^{2} = \int \|\h_m(\nu)\|^{2} \drm \nu$. Condition \eqref{assumption:Hnu} thus implies that the SNR 
$\frac{\mathbb{E}|u_{m,n}|^{2}}{\mathbb{E}|v_{m,n}|^{2}}$ on each sensor $m$ converges towards $0$ when $M \rightarrow +\infty$. 
This condition in particular holds if the useful signal powers received on the various sensors are of the same order of magnitude, in which case
$\frac{\mathbb{E}|u_{m,n}|^{2}}{\mathbb{E}|v_{m,n}|^{2}} = \Ocal(\frac{1}{M})$.

\section{Asymptotic behaviour of the SCM $\hat{\C}_{\y}(\nu)$}

In this section, we study the SCM $\hat{\C}_{\y}(\nu)$ under the asymptotic regime \eqref{assumption:regime}, and show that this matrix can be approximated by a standard model from RMT. To that purpose, we study separately the pure noise case ($\y_n = \v_n$ in \eqref{definition:model})
and the noise free case ($\y_n = \u_n$ in \eqref{definition:model}), and in the following, we denote by
\begin{align}
	\xibs_{\v}(\nu) = \frac{1}{\sqrt{N}}\sum_{n=1}^N \v_n \erm^{-\irm 2 \pi \nu (n-1)}
	\notag\\
	\xibs_{\u}(\nu) = \frac{1}{\sqrt{N}}\sum_{n=1}^N \u_n \erm^{-\irm 2 \pi \nu (n-1)}
	\notag
\end{align}
the finite Fourier transforms of $(\v_n)_{n \in\Zbb}$ and $(\u_n)_{n \in \Zbb}$ respectively, and also use the notation $\Vcal_N = \{0, \frac{1}{N},\ldots,\frac{N-1}{N}\}$ for the set of Fourier frequencies. Due to space constraints, the proofs of the results are omitted.

We have the following approximation result regarding the finite Fourier transforms of $(\v_n)_{n \in\Zbb}$.
\begin{theorem}
	\label{theorem:noise}
	Let $\Sigmabs_{\v}(\nu) = \frac{1}{\sqrt{B+1}}[\xibs_{\v}\left(\nu-\frac{B}{2N}\right),\ldots,\xibs_{\v}\left(\nu+\frac{B}{2N}\right)]$.
	Then under Assumption \ref{assumption:noise}, for all $\nu \in \Vcal_N$, there exists a $M \times (B+1)$ random matrix $\Z(\nu)$ with i.i.d.
	$\Ncal_{\Cbb}(0,1)$ entries such that
	\begin{align}
		\max_{\nu \in \Vcal_N} \left\|\Sigmabs_{\v}(\nu) - \frac{1}{\sqrt{B+1}} \S_{\v}(\nu)^{\frac{1}{2}} \Z(\nu)\right\|
		\xrightarrow[N\to\infty]{a.s.} 0
		\notag
	\end{align}
\end{theorem}
Using standard concentration bound on the norm of Wishart matrices (see e.g. \cite{Vershynin2012}), we deduce that 
$\max_{\nu \in \Vcal_N}\|\Z(\nu)\| = \Ocal(\sqrt{B})$ a.s. and
\begin{align}
	\max_{\nu \in \Vcal_N}\left\|\hat{\S}_{\v}(\nu) - \frac{1}{B+1} \S_{\v}(\nu)^{\frac{1}{2}} \Z(\nu)\Z(\nu)^*\S_{\v}(\nu)^{\frac{1}{2}} \right\|
	\xrightarrow[N\to\infty]{a.s.} 0
	\label{conv_Sv}
\end{align}
where $$\hat{\S}_{\v}(\nu) = \frac{1}{B+1} \sum_{b=-B/2}^{B/2} \xibs_{\v}(\nu+b/N)\xibs_{\v}(\nu+b/N)^*$$ 
Thus, Theorem \ref{theorem:noise} shows that the smoothed periodogram of $(\v_n)_{n \in \Zbb}$ behaves (asymptotically in operator norm) as a Wishart matrix with scale matrix coinciding with the spectral density $\S_{\v}(\nu)$. 

We now turn to the study of the finite Fourier transforms of the signal part $(\u_n)_{n \in\Zbb}$.
\begin{theorem}
	\label{theorem:signal}
	Let $\Sigmabs_{\u}(\nu) = \frac{1}{\sqrt{B+1}}[\xibs_{\u}\left(\nu-\frac{B}{2N}\right),\ldots,\xibs_{\u}\left(\nu+\frac{B}{2N}\right)]$ and 
	$\Sigmabs_{\epsilonbs}(\nu) = \frac{1}{\sqrt{B+1}}[\xibs_{\epsilonbs}\left(\nu-\frac{B}{2N}\right),\ldots,\xibs_{\epsilonbs}\left(\nu+\frac{B}{2N}\right)]$.
	Then under Assumption \ref{assumption:signal}, it holds that
	\begin{align}
		\max_{\nu \in \Vcal_N}\left\|\Sigmabs_{\u}(\nu) - \H(\nu)\Sigmabs_{\epsilonbs}(\nu)\right\|
		\xrightarrow[N\to\infty]{a.s.} 0
		\notag
	\end{align}
\end{theorem}
Note that the type of approximation stated in Theorem \ref{theorem:signal} is well-known in the classical large sample size regime 
in which $M,K,B$ are fixed while $N \to \infty$ since in that case \cite[Th. 4.5.2]{Brillinger1981}
\begin{align}
	\sup_{\nu \in [0,1]} \left\|\Sigmabs_{\u}(\nu) - \H(\nu)\Sigmabs_{\epsilonbs}(\nu)\right\| = \Ocal\left(\sqrt{\frac{\log(N)}{N}}\right)
	\quad a.s.
	\notag
\end{align}
Of course, in the high dimensional regime in which $M$ and $B$ also converge to infinity, the result of Theorem \ref{theorem:signal} cannot be deduced from
\cite{Brillinger1981} and requires a specific study.

Using Theorems \ref{theorem:noise} and \ref{theorem:signal}, we directly obtain as for \eqref{conv_Sv}, that for all $\nu \in \Vcal_N$ there exists
a $M \times (B+1)$ random matrix $\X(\nu)$ with i.i.d. $\Ncal_{\Cbb}(0,1)$ entries such that 
\begin{align}
		\max_{\nu \in \Vcal_N}
		\left\|\hat{\S}_{\y}(\nu) - \frac{1}{B+1}\S_{\y}(\nu)^{\frac{1}{2}}\X(\nu)\X(\nu)^*\S_{\y}(\nu)^{\frac{1}{2}}\right\| \xrightarrow[N\to\infty]{a.s.} 0
		\label{conv_Sy}
\end{align}
Moreover, using Assumption \ref{assumption:signal} (condition \eqref{assumption:Hnu}), we can show that 
\begin{align}
	\max_{\nu \in  \Vcal_N}\left\|\diag\left(\hat{\S}_{\y}(\nu)\right) - \S_{\v}(\nu)\right\| \xrightarrow[N\to\infty]{a.s.} 0
	\label{conv_diagSy}
\end{align}
Equipped with \eqref{conv_Sy} and \eqref{conv_diagSy}, we are now in position to study the behaviour of the SCM $\hat{\C}_{\y}(\nu)$.
\begin{corollary}
	\label{corollary:SCM}
	Under Assumptions \ref{assumption:noise} and \ref{assumption:signal}, there exists a $M \times (B+1)$ random matrix $\X(\nu)$ with i.i.d.
	$\Ncal_{\Cbb}(0,1)$ entries such that
	\begin{align}
		\max_{\nu \in \Vcal_N}
		\left\|\hat{\C}_{\y}(\nu) - \Xibs(\nu)^{\frac{1}{2}}\frac{\X(\nu)\X(\nu)^*}{B+1}\Xibs(\nu)^{\frac{1}{2}}\right\| \xrightarrow[N\to\infty]{a.s.} 0
		\label{conv_SCM_norm}
	\end{align}
	where $\Xibs(\nu) = \S_{\v}(\nu)^{-\frac{1}{2}}\H(\nu)\H(\nu)^*\S_{\v}(\nu)^{-\frac{1}{2}} + \I$.
\end{corollary}
Let us make two observations regarding Corollary \ref{corollary:SCM}.

First, notice that the operator norm approximation \eqref{conv_SCM_norm} implies, thanks to Weyl's inequality, that
\begin{align}
	&\max_{m=1,\ldots,M} \max_{\nu \in \Vcal_N}
	\Bigl|
		\lambda_m\left(\hat{\C}_{\y}(\nu)\right) 
		-
		\notag\\ 
	& \qquad\qquad\lambda_m\left(\Xibs(\nu)^{\frac{1}{2}}\frac{\X(\nu)\X(\nu)^*}{B+1}\Xibs(\nu)^{\frac{1}{2}}\right)
	\Bigr|
	\xrightarrow[N\to\infty]{a.s.} 0
	\label{conv_SCM_eig}
\end{align}
which shows that any linear spectral statistic \cite{Bai2004} based on the spectrum of $\hat{\C}_{\y}(\nu)$ behaves (at first order) as the same linear spectral statistic applied to the spectrum of the corresponding Wishart matrix $\Xibs(\nu)^{\frac{1}{2}}\frac{\X(\nu)\X(\nu)^*}{B+1}\Xibs(\nu)^{\frac{1}{2}}$. 

Second, an interpretation of the results of Theorems \ref{theorem:noise} and \ref{theorem:signal}, and Corollary \ref{corollary:SCM} can be given in the light of array processing application. Indeed, the time domain model \eqref{definition:model} is usually referred to as wideband model, and in particular the useful signal contribution $(\u_n)_{n \in \Zbb}$ is not necessarily confined to a low-dimensional subspace of $\Cbb^M$ due to the filtering induced by $(\H_k)_{k \geq 0}$. In the frequency domain, Theorems \ref{theorem:noise} and \ref{theorem:signal} show that we retrieve in some sense a narrowband model, and that the useful signal contribution is now confined to $K$-dimensional subspace of $\Cbb^M$, which opens the possibility to use standard narrowband techniques for e.g. detecting the presence of the useful signal $(\u_n)_{n \in \Zbb}$, using test statistics based on the eigenvalues of $\hat{\S}_{\y}(\nu)$ or $\hat{\C}_{\y}(\nu)$ at multiple Fourier frequencies $\nu \in \Vcal_N$. 

We take profit of these two observations to study the consistency of a certain spectral detection test in the next section.

\section{Application to spectral detection}

As we have seen in the previous section, the SCM $\hat{\C}_{\y}(\nu)$ behaves asymptotically in operator norm as a Wishart matrix, with associated scale matrix 
$\Xibs(\nu) = \S_{\v}(\nu)^{-\frac{1}{2}}\H(\nu)\H(\nu)^*\S_{\v}(\nu)^{-\frac{1}{2}} + \I$ being a fixed rank $K$ perturbation of the identity matrix. 
We can therefore exploit the well-known results on spiked models in the RMT literature \cite{Baik2006}. We define for the remainder the function
\begin{align}
%	\phi(\lambda) = 
%	\begin{cases}	
%		\lambda + \frac{c \lambda}{\lambda - 1} & \text{ if } \lambda > 1+\sqrt{c}
%		\\
%		\left(1+\sqrt{c}\right)^2 & \text{ if } \lambda \leq 1+\sqrt{c}
%	\end{cases}.
	\phi(\gamma) = 
	\begin{cases}	
		\frac{(\gamma+1)(\gamma+c)}{\gamma} & \text{ if } \gamma > \sqrt{c}
		\\
		\left(1+\sqrt{c}\right)^2 & \text{ if } \gamma \leq \sqrt{c}
	\end{cases}
	\notag
\end{align}
and fix a frequency $\nu_N^* \in \Vcal_N$ such that
\begin{align}
	\nu_N^* \in \argmax_{\nu \in \Vcal_N} \lambda_1\left(\S_{\v}(\nu)^{-\frac{1}{2}}\H(\nu)\H(\nu)^*\S_{\v}(\nu)^{-\frac{1}{2}}\right)
	\notag
\end{align}
We consider the following purely technical additional assumption.
\begin{assumption}
	\label{assumption:spike}
%	For all $k \in \{1,\ldots,K\}$, there exists a function $\gamma_k:[0,1] \to (0,\infty)$ such that
%	\begin{align}
%		\sup_{\nu \in [0,1]} 
%		\left|
%			\lambda_k\left(\S_{\v}(\nu)^{-\frac{1}{2}}\H(\nu)\H(\nu)^*\S_{\v}(\nu)^{-\frac{1}{2}}\right)
%			-
%			\gamma_k(\nu)
%		\right|
%		\xrightarrow[N\to\infty]{} 0.
%		\notag
%	\end{align}	
	For all $k\in \{1,\ldots,K\}$, there exists $\gamma_k > 0$ such that
	\begin{align}
		\lambda_k\left(\S_{\v}(\nu_N^*)^{-\frac{1}{2}}\H(\nu_N^*)\H(\nu_N^*)^*\S_{\v}(\nu_N^*)^{-\frac{1}{2}}\right) \xrightarrow[N\to\infty]{} \gamma_k
		\notag
	\end{align}
\end{assumption}
From \eqref{conv_SCM_eig} and the results of \cite{Baik2006}, we deduce the following indivual behaviour of the eigenvalues of the SCM.
\begin{corollary}
	\label{corollary:spike}
	Under Assumptions \ref{assumption:noise}, \ref{assumption:signal} and \ref{assumption:spike}, for all $k=1,\ldots,K$,
	\begin{align}
		\lambda_{k}\left(\hat{\C}_{\y}(\nu_N^*)\right) \xrightarrow[N\to\infty]{a.s.} \phi\left(\gamma_k\right)
		\notag
	\end{align}
	whereas 
	\begin{align}
		\lambda_{K+1}\left(\hat{\C}_{\y}(\nu_N^*)\right) &\xrightarrow[N\to\infty]{a.s.} \left(1+\sqrt{c}\right)^2
		\notag \\
		\lambda_{M}\left(\hat{\C}_{\y}(\nu_N^*)\right) &\xrightarrow[N\to\infty]{a.s.} \left(1-\sqrt{c}\right)^2
		\notag
	\end{align}	
\end{corollary}
Consider the set of indexes $\Kcal = \{k \in \{1,\ldots,K\} : \gamma_k > \sqrt{c}\}$.
Corollary \ref{corollary:spike} implies that each "signal" eigenvalue $\lambda_k(\hat{\C}_{\y}(\nu_N^*))$ of the SCM at frequency $\nu_N^*$ for which $k \in \Kcal$ asymptotically splits from the "noise" eigenvalues $\lambda_{K+1}(\hat{\C}_{\y}(\nu_N^*)),\ldots,\lambda_{M}(\hat{\C}_{\y}(\nu_N^*))$, which concentrate in a neighborhood of the interval $[(1-\sqrt{c})^2,(1+\sqrt{c})^2]$. Likewise, the signal eigenvalues $\lambda_k(\hat{\C}_{\y}(\nu_N^*))$ for which $k \not\in \Kcal$
are asymptotically absorbed in a neighborhood of $[(1-\sqrt{c})^2,(1+\sqrt{c})^2]$. This phase transition phenomenon for the $k$-th eigenvalue of $\hat{\C}_{\y}(\nu_N^*)$ thus occurs when the eigenvalues of $\S_{\v}(\nu_N^*)^{-\frac{1}{2}}\H(\nu_N^*)\H(\nu_N^*)^*\S_{\v}(\nu_N^*)^{-\frac{1}{2}}$ are sufficiently large:
\begin{align}
	\gamma_k = \lim_{N\to\infty}\lambda_k\left(\S_{\v}(\nu_N^*)^{-\frac{1}{2}}\H(\nu_N^*)\H(\nu_N^*)^*\S_{\v}(\nu_N^*)^{-\frac{1}{2}}\right) > \sqrt{c}
	\notag
\end{align}

The result of Corollary \ref{corollary:spike} can be exploited to obtain consistent test statistics based on the eigenvalues of $\hat{\C}_{\y}$, 
for detecting the presence of the useful signal in model \eqref{definition:model}, i.e. considering the hypothesis test
\begin{align}
	\Hcal_0: \y_n = \v_n \quad\text{vs}\quad \Hcal_1:\y_n=\u_n+\v_n
	\notag
\end{align}
Since the intrinsic dimensionality $K$ of the useful signal is not necessarily known in practice, we consider the detection test solely based on the largest eigenvalue of the SCM, i.e.
\begin{align}
	T_{\epsilon} = \mathbbm{1}_{\left((1+\sqrt{c})^2+\epsilon,+\infty\right)}\left(\max_{\nu \in \Vcal_N}\left\|\hat{\C}_{\y}(\nu)\right\|\right)
	\notag
\end{align}
where $\epsilon$ is some threshold. We thus have the following consistency result.
\begin{theorem}
	Under Assumptions \ref{assumption:noise}, \ref{assumption:signal} and \ref{assumption:spike}, and if 
	\begin{align}
		\gamma_1 > \sqrt{c}
		\notag
	\end{align}		
	then for all $\epsilon \in \left(0, \phi(\gamma_1) - (1+\sqrt{c})^2\right)$, and $i \in\{0,1\}$
	\begin{align}
		\Pbb_i\left(\lim_{N\to\infty}T_{\epsilon} = i\right) = 1
		\notag
	\end{align}
	where $\Pbb_i$ is the underlying probability measure under hypothesis $\Hcal_i$.
\end{theorem}
In practical situations, the condition $\gamma_1 > \sqrt{c}$ means that the quantity
$\|\S_{\v}(\nu^*_N)^{-1}\H(\nu^*_N)\H(\nu^*_N)^*\|$ should be larger than $\sqrt{\frac{M}{B}}$.

We now illustrate numerically the above asymptotic approximations when $K=1$. 
The noise is chosen as a MA(1) process with standard Gaussian innovation $(\z_n)_{n \in \Zbb}$, i.e. $\v_n = \z_n + \theta_1 \z_{n-1}$, 
whereas for the useful signal, we choose 
%\begin{align}
	$\H_k=C_{\rm SNR} \frac{1}{\sqrt{M}} \beta^k \left(1,\ldots,1\right)^T$
%	\notag
%\end{align}
with $\beta = \frac{10}{11}$, and $C_{\rm SNR}$ a factor controlling the SNR defined below. 
%We also approximate $\u_n$ by truncation by setting $\u_n= \sum_{k=0}^{100} \H_k\epsilon_{n-k}$. 
%$C_{SNR}$ is a constant controlling the intensity of the signal.

\begin{figure}[h!]
	\centering
	\includegraphics[scale=0.25]{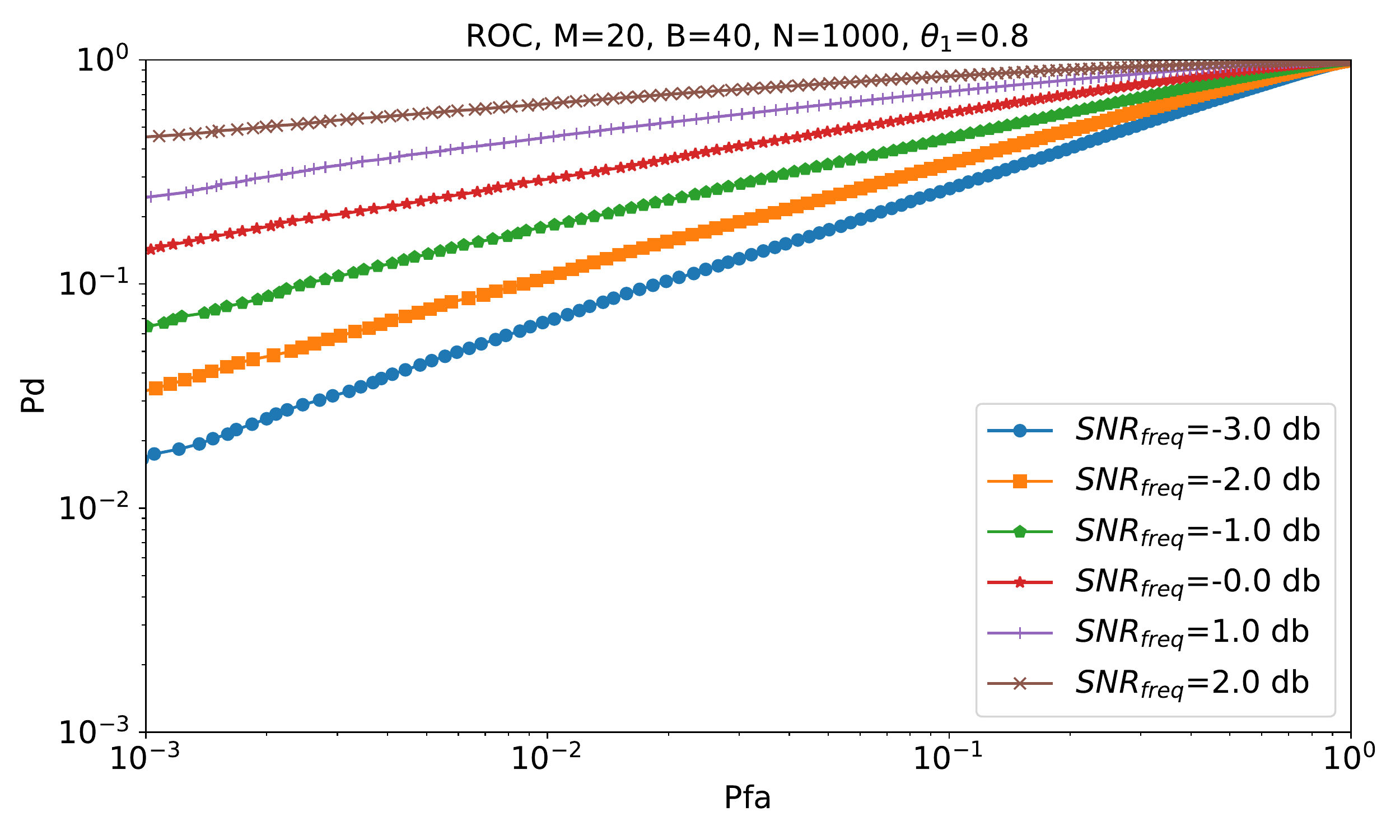}
	\caption{ROC curve of $T_{\epsilon}$ with varying SNR}
	\label{fig2}
\end{figure}

\begin{figure}[h!]
	\centering
	\includegraphics[scale=0.25]{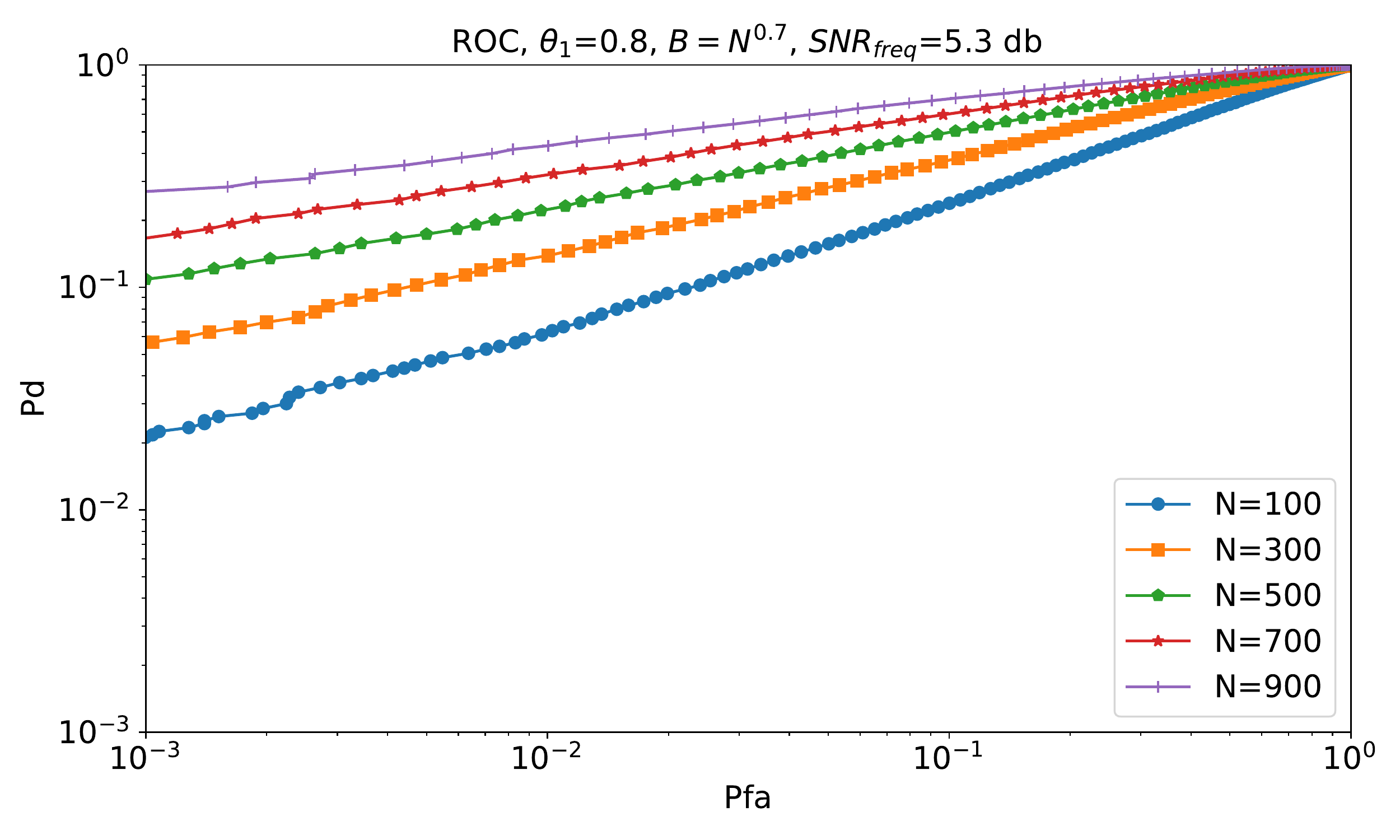}
	\caption{ROC curve of $T_{\epsilon}$ with varying $N$}
	\label{fig3}
\end{figure}

Figures 
%\ref{fig1}, 
\ref{fig2} and \ref{fig3} represent the probability of detection as a function of the probability of false alarm (ROC curve) for the test statistic $T_{\epsilon}$, for different values of the threshold $\epsilon$ and a total of $10^5$ draws. 

%In Figure \ref{fig1}, we set $B=40$ and $M=20$, with $N$ varying from $100$ to $1900$. We notice that as the ratio $B/N$ decreases, the performance of the test improves. 
%\begin{figure}
%	\centering	
%	\includegraphics[scale=0.27]{./ROC_BsurN.pdf}
%	\caption{ROC curve of $T_{\epsilon}$ with varying ratio $\frac{B}{N}$}
%	\label{fig1}
%\end{figure}
In Figure \ref{fig2}, the ROC curve is plotted for different values of SNR defined in the frequency domain as follows
\begin{align}
	\mathrm{SNR}_{\rm freq} = \sup_{\nu \in [0,1]} \sum_{m=1}^M \frac{\|\h_m(\nu)\|^2}{s_m(\nu)}
	\notag
\end{align}
As the SNR increases, the performance of the test improves as well. 
In Figure \ref{fig3}, we compute the ROC curve for increasing values of $N$ with $B=N^{0.7}$ and $B=2M$. As $N$ grows, the performance of the test improves as expected since we get closer to the high dimensional asymptotic regime \eqref{assumption:regime}.

%\section{Conclusion}
%\section{Simulations}

%\section{References}
%
\newpage
\bibliographystyle{IEEEbib}
\bibliography{icassp20}

\end{document}